\documentclass[twocolumn,aps,prb,showpacs]{revtex4}
\usepackage{epsfig}
\usepackage{graphicx}
\usepackage{color}

\begin{document}

\title{Eliashberg theory of superconductivity and inelastic rare-earth impurity scattering in filled skutterudite
La$_{1-x}$Pr$_{x}$Os$_{4}$Sb$_{12}$}

\author{Jun Chang$^1$, Ilya Eremin$^{1,2}$, Peter Thalmeier$^3$, and Peter Fulde$^{1}$}

\affiliation{$^1$ Max-Planck-Institut f\"ur Physik komplexer
Systeme, D-01187 Dresden, Germany \\
$^2$ Institute f\"ur Mathematische und Theoretische  Physik,
TU-Braunschweig,
D-38106 Braunschweig, Germany \\
$^3$ Max-Planck-Institut f\"ur Chemische Physik fester Stoffe,
D-01187 Dresden, Germany}
\date{\today}

\begin{abstract}
We study the influence of inelastic rare-earth impurity scattering
on electron-phonon-mediated superconductivity and mass
renormalization in La$_{1-x}$Pr$_{x}$Os$_{4}$Sb$_{12}$ compounds.
Solving the strong-coupling Eliashberg equations we find that the
dominant quadrupolar component of the inelastic scattering on Pr
impurities yields an enhancement of the superconducting transition
temperature $T_c$ in LaOs$_{4}$Sb$_{12}$ and increases monotonically
as a function of Pr concentration. The calculated results are in
good agreement with the experimentally observed $T_c(x)$
dependence. Our analysis suggests that phonons and quadrupolar
excitations cause an attractive electron interaction which results
in the formation of Cooper pairs and singlet superconductivity in
PrOs$_{4}$Sb$_{12}$.
\end{abstract}
\maketitle


The Pr-based filled-skutterudite compounds have attracted much
attention because of their exotic properties like metal-insulator
transition or unusual heavy-fermion
behavior.\cite{sekine,sato,sugawara} This also concerns recently
discovered PrOs$_{4}$Sb$_{12}$, the first Pr-based heavy-fermion
superconductor with $T_c=1.85$ K which possesses many exotic
properties compared to that of the Ce- and U-based
superconductors.\cite{Bauer} The heavy-electron mass has been
confirmed by the large specific heat jump $\Delta C / T_c \sim 500
\ mJ/(K^2\ mol)$ at $T_c$ (see Ref.~\onlinecite{Bauer}) and by de
Haas-van Alphen (dHvA) measurements.\cite{sugawaraF} The ground
state in the crystalline electric field (CEF) for Pr$^{3+}$ ion is a
$\Gamma_1$ singlet, which is separated by the first excited state, a
$\Gamma_4^{(2)}$ triplet by a gap of $\Delta_{CEF} \sim
8$ K.\cite{CEF} Because of the small $\Delta_{CEF}$, the relation
between the quadrupole fluctuations associated with the
$\Gamma_4^{(2)}$ state and the superconductivity has been recently
the focus of intense discussions.\cite{quadr1,quadr2}

At present, the Cooper-pairing mechanism and the corresponding
symmetry of the superconducting gap in PrOs$_{4}$Sb$_{12}$ is still
under debate. For example, initial studies of the thermal
conductivity in a rotated magnetic field have suggested the presence
of two distinct SC phases.\cite{Izawa} Similarly, the London
penetration depth \cite{Huxley} has indicated a possible nodal
structure of the superconducting gap. On the contrary, a number of
experimental techniques such as nuclear-quadrupole-resonance (NQR),
\cite{Yogi,Kotegawa} Scanning Tunneling Microscopy(STM),
\cite{Suderow} muon spin relaxation, ($\mu SR$) \cite{MacLaughlin}
thermal conductivity,\cite{Seyfarth} and specific heat
measurements\cite{sakakibara} are in agreement with a fully
developed isotropic $s$-wave-like superconducting gap at the Fermi
surface (FS). Another interesting experimental observation by means
of the dHvA effect was that PrOs$_{4}$Sb$_{12}$ has a very similar FS
topology as the conventional $s$-wave superconductor
LaOs$_{4}$Sb$_{12}$ (Ref.~\onlinecite{sugawaraF}) ($T_c=0.74$ K) which hints at a
relatively weak hybridization of the conduction band with Pr$^{3+}$
4$f^2$ electrons. Furthermore, it has been found\cite{Yogi,Rotundu}
that in the alloy compound La$_{1-x}$Pr$_{x}$Os$_{4}$Sb$_{12}$ the
superconducting transition temperature changes smoothly upon changing
$x$. These observations raise doubts about the unconventional nature of
the Cooper pairing in PrOs$_{4}$Sb$_{12}$. But the enhancement of
$T_c$ and the effective quasiparticle mass in this compound with
respect to that of LaOs$_{4}$Sb$_{12}$ have to be understood.

It is well known that scattering by magnetic impurities suppresses
conventional $s$-wave superconductivity by destroying Cooper pairs in
a singlet state. This situation, however, can change in the case of
paramagnetic non-Kramers rare-earth impurities. For example, it has
been shown previously\cite{Fulde} that for superconductors
containing an impurity with crystal-field-split energy levels the
inelastic charge scattering of conduction electrons from the
aspherical part of the 4$f$ charge distribution may yield an
increase of $T_c$. However, in most of the cases the usual magnetic
exchange interaction is dominant, thus suppressing $T_c$. In this Rapid 
Communication, we solve the nonlinear Eliashberg equations for
electron-phonon mediated Cooper pairing including inelastic
scattering on Pr impurities. We find that a dominant quadrupolar
scattering introduced by the Pr ions is responsible for an increase
of $T_c$ in La$_{1-x}$Pr$_{x}$Os$_{4}$Sb$_{12}$ as a function of
Pr concentration.

\begin{figure}[t]
\includegraphics[width=0.6\columnwidth]{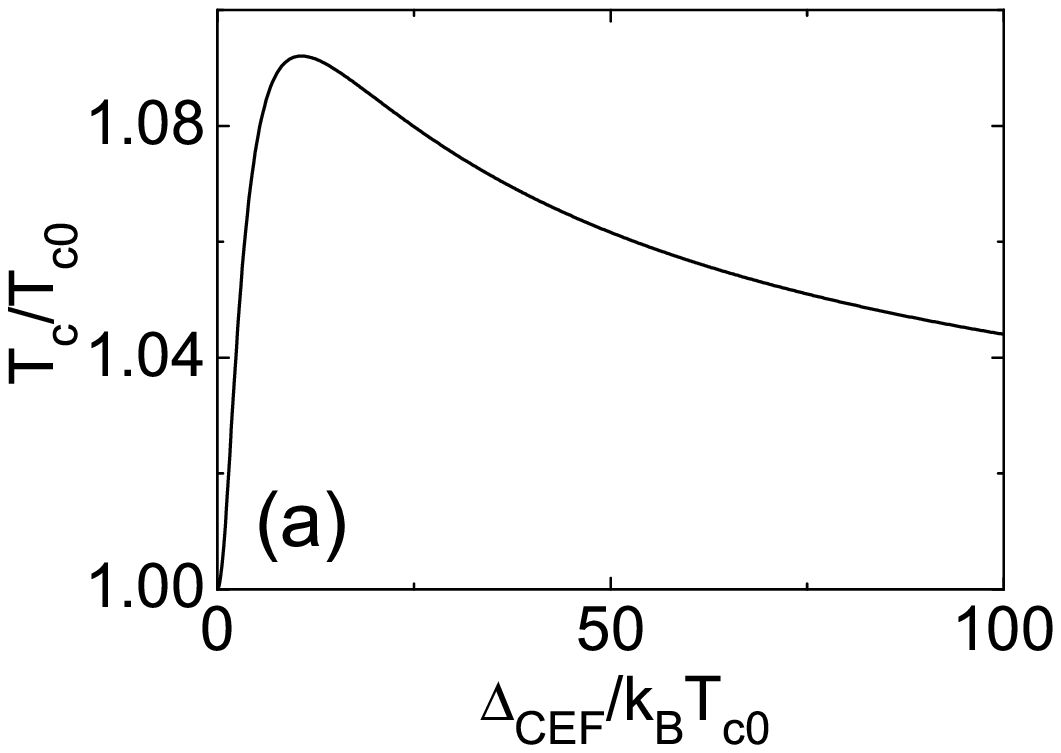}
\includegraphics[width=0.6\columnwidth]{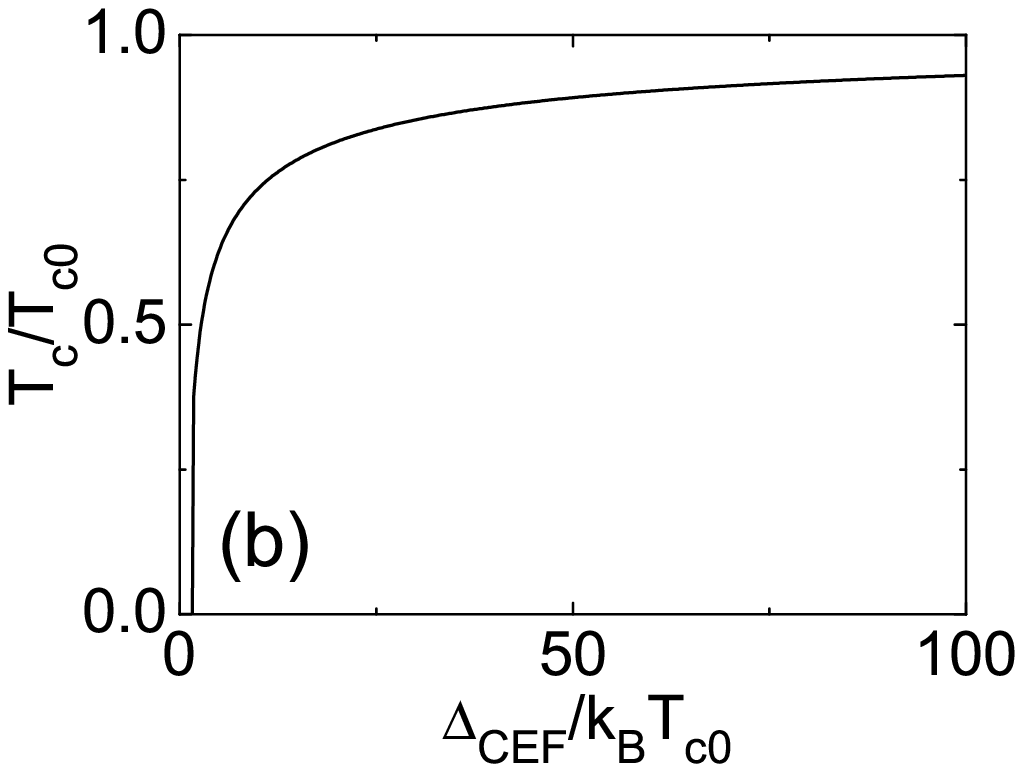}
\caption{Calculated values of the superconducting transition
temperatures as a function of $\Delta_{CEF}/T_{c0}$ for inelastic
quadrupolar (a) and magnetic (b) scattering by impurities with two
singlet levels using Eqs. (1).} \label{fig1}
\end{figure}

Before solving the nonlinear version of the Eliashberg equations
let us illustrate the results of Ref.~\onlinecite{Fulde} where the
influence of inelastic scattering by rare-earth impurities on the
superconducting transition temperature $T_{c0}$ without impurities
has been studied for both exchange and quadrupolar scattering. In
particular, for impurities with two singlet levels separated by an
energy $\Delta_{CEF}$ the analog of the Abrikosov-Gor'kov relation
for the change of $T_c$ as a function of the total impurity
scattering rate \cite{Fulde} for the $s$-wave superconductor is
given by
\begin{eqnarray}
-\frac{8}{\pi} T_c \tau_{12}^M \ln \frac{T_c}{T_{c0}} & = &
1+\frac{\tanh x}{x} -\left(\tanh x \right)^2 -A(x) +B(x), \nonumber\\
-\frac{8}{\pi} T_c \tau^Q_{12} \ln \frac{T_c}{T_{c0}} & = &
1-\frac{\tanh x}{x} -\left(\tanh x \right)^2 + A(x).
\label{eq1}
\end{eqnarray}
Here, $1/\tau_{12}^M$ and $1/\tau_{12}^Q$ are the magnetic and
quadrupolar scattering rate, respectively, and
$x=\Delta_{CEF}/2 k_B T_{c0}$. Furthermore, $A(x)$ and
$B(x)$ are the combinations of digamma functions\cite{Fulde} and are
smooth functions of $x$. In Figs. \ref{fig1}(a) and \ref{fig1}(b) we show the
behavior of $T_c/ T_{c0}$ as a function of
$\Delta_{CEF}/k_B T_{c0}$ for the quadrupolar (charge) and
exchange (magnetic) scattering, respectively. Assuming equal matrix
elements for both scatterings ($\tau_{12}^M T_{c0} = \tau_{12}^Q
T_{c0} =\tau_{12} T_{c0}$) and setting somewhat arbitrarily
$\tau_{12} T_{c0}= 1.4$ we find very different behavior. While for
quadrupolar scattering $T_c$ increases with respect to T$_{c0}$,
with a pronounced maximum around $\Delta_{CEF}/k_B T_{c0} \approx
10$, the magnetic scattering results in a complete suppression of
$T_c$ at small $\Delta_{CEF}$ and gives a moderate reduction at
larger $x$ values. Overall one finds that for similar scattering
amplitudes the magnetic exchange scattering dominates and $T_c$
decreases.

Let us consider more specifically Pr$^{3+}$ impurities in
La$_{1-x}$Pr$_{x}$Os$_{4}$Sb$_{12}$ compounds. As mentioned
above the CEF level scheme is composed of a $\Gamma_1$ singlet
ground state, a $\Gamma_4^{(2)}$ triplet first excited state
with $\Delta_{CEF} \approx 8$ K, and other excited
states located at much larger energies above $\sim 100$ K. \cite{CEF}
Furthermore, due to $T_h$ crystal group symmetry, the
$\Gamma_4^{(2)}$ level is of the form
$|\Gamma_4^{(2)}(m) \rangle = \sqrt{1-d^2}|\Gamma_5 (m) \rangle + d |\Gamma_4
(m)\rangle,$
where $m=+,-,0$. Here, $\Gamma_5$ and $\Gamma_4$ are the wave
functions for the cubic $O_h$ symmetry and $d$ is a parameter that
describes the mixture of the two states and is estimated to be equal 0.26.
\cite{Shiina} Since scattering between $\Gamma_1$ and $\Gamma_5$ is
quadrupolar, the dominant scattering between $\Gamma_{4}^{(2)}$ and
$\Gamma_{1}$ will be quadrupolar as well. Furthermore, from
Fig.~\ref{fig1}(a) we see that the enhancement of $T_c$ is largest
for $\Delta_{CEF} \approx 8$ K. Keeping in mind that the higher
excited states lie at energies $\geq 100$ K the magnetic
pair-breaking scattering from these levels will play only a minor role in
affecting $T_c$. This suggests that Pr$^{3+}$ions in
La$_{1-x}$Pr$_{x}$Os$_{4}$Sb$_{12}$ yield a strong Cooper-pair
enhancement as compared with LaOs$_{4}$Sb$_{12}$. At the same time,
the Pr$^{3+}$ ions in La$_{1-x}$Pr$_{x}$Os$_{4}$Sb$_{12}$ can be
treated as independent impurities only at low enough doping. At
large Pr concentrations the RKKY interaction between the 4$f^2$ ions
results in the formation of $O_{xy}$-type quadrupolar exciton
bands as confirmed experimentally.\cite{quadr2} The latter can be
treated as low-energy bosonic excitations which contribute in addition to
phonons to the Cooper-pairing. But one should notice that
the exciton dispersion is relatively weak, {\it i.e.}, about $10\%$ of
$\Delta_{CEF}$.\cite{quadr2}

Using the generalized Holstein-Primakoff (HP) method,\cite{Shiina2003} the low-energy Hamiltonian describing the magnetic and quadrupolar interaction between conduction electrons
and the CEF split energy levels of Pr$^{3+}$ is given by
%
\begin{eqnarray}
\lefteqn{H = \sum_{{\bf q}u} \omega_{{\bf q}} \beta_{{\bf
q}u}^{\dagger}\beta_{{\bf q}u} - I_{ac} \sum_{{\bf k,q},\sigma,u}
f_{u}({\bf q}) \lambda^{Q}_{\bf q} \hat{O}_{{\bf q}u} c_{{\bf k}
\sigma}^{\dagger} c_{{\bf k+q} \sigma}} && \nonumber\\
&& - (1-g_{L})I_{ex} \sum_{{\bf k,q} ,\sigma,\sigma',u}
\sigma_{\sigma\sigma'}^{u}\lambda^M_{\bf q} \hat{J}_{{\bf q}u}
c_{{\bf k}\sigma}^{\dagger}c_{{\bf k+q}\sigma'}, \label{exciton}
\end{eqnarray}
%
where $\omega_{{\bf q}}$ is the energy dispersion of the exciton.
There are three bosonic modes ($u=a,b,c$) corresponding to the
excitations between singlet $\Gamma_{1}$ and triplet
$\Gamma_{4}^{(2)}$ states. $\sigma^{u}$ are the Pauli matrices
($c=x$, $b=z$, $a=y$) and $f_{c}({\bf q})=\hat{q}_{x} \hat{q}_{y}$,
$f_{b}({\bf q})=\hat{q}_{z}\hat{q}_{x}$, and $f_{a}({\bf
q})=\hat{q}_{y}\hat{q}_{z}$ are the form factors of the quadrupolar
interaction. The quadrupolar and magnetic excitations are written in
terms of the bosonic operators $\hat{O}_{{\bf
q}u}=i\left(\beta_{{\bf q}u}-\beta_{-{\bf q}u}^{\dagger}\right)$ and
$\hat{J}_{{\bf q}u}=\left(\beta_{{\bf q}u}+\beta_{-{\bf
q}u}^{\dagger}\right)$ with $ \left( \lambda^Q_{\bf q} \right) ^{2}
= (\Delta_{CEF}+2D_{M}z \gamma_{\bf q})/\omega_{\bf q}$ and
$\left(\lambda^M_{ \bf q} \right) ^{2} =
(\Delta_{CEF}+2D_{Q}z\gamma_{\bf q})/ \omega_{\bf q}$,
respectively. Here, $D_M$ and $D_Q$ are the effective magnetic and
quadrupolar coupling constants,\cite{Shiina} $z$ is the coordination
number, and $\gamma_{\bf k}$ is the corresponding structure factor.
Due to the weak dispersion of the excitons, $\lambda^Q_{\bf q}$ and
$\lambda^M_{\bf q}$ are almost momentum independent. Solving the
Eliashberg equations, we use an effective interaction between
quadrupolar excitons and conduction electrons averaged over the FS.

On the real frequency axis the finite temperature Eliashberg equations for the superconducting
gap $\Delta (\omega, T)$ and the renormalization function $Z(\omega,
T)$  are given by \cite{holcomb}
\begin{eqnarray}
\Delta(\omega,T) &=& \frac{1}{Z(\omega,T)} \int_{0}^{\infty}
d\omega' \mbox{Re} \left\{
\frac{\Delta(\omega',T)}{\sqrt{\omega'^{2}-\Delta^{2}(\omega',T)}}\right\}\nonumber\\
&\times&\left[K_{+}(\omega,\omega',T)-\mu^{*} \tanh
\left(\frac{\beta\omega'}{2} \right) \right],
\end{eqnarray}
\begin{eqnarray}
\omega(1-Z(\omega,T)) &=& \int_{0}^{\infty} d\omega' \mbox{Re}
\left\{
\frac{\omega'}{\sqrt{\omega'^{2}-\Delta^{2}(\omega',T)}} \right\}\nonumber\\
&\times&K_{-} (\omega,\omega',T),
\end{eqnarray}
where
\begin{eqnarray}
K_{\pm}(\omega,\omega',T) = &&\int_{0}^{\infty} d\Omega
H_{\pm}(\Omega)\nonumber\\
&&\times \left[ \frac{f(-\omega')+n(\Omega)}{\omega' + \omega + \Omega} 
  \pm \frac{f(-\omega')+n(\Omega)}{\omega' - \omega +
\Omega}  \right.\nonumber\\
&&\left.\mp \frac{f(\omega')+n(\Omega)}{-\omega' + \omega + \Omega}
- \frac{f(\omega')+n(\Omega)}{-\omega' - \omega + \Omega} \right],
\end{eqnarray}
and
$H_{\pm}(\Omega) = \alpha_{P}^{2}G(\Omega) + \alpha_{Q}^{2}F(\Omega)
\mp \alpha_{M}^{2}F(\Omega)$
is the generalized electron-boson coupling function averaged over
the FS. Here, $\mu^*$ is the screened Coulomb repulsion, $f(\omega)$
and $n(\omega)$ are the Fermi and Bose distribution functions,
respectively, and $\beta = 1 / k_B T$.

As mentioned above there are two types of bosonic excitations. In
LaOs$_{4}$Sb$_{12}$ the superconductivity is driven by the
electron-phonon interaction and we assume a single Lorentzian phonon
mode at $\hbar \omega_{E} = 26$ meV with broadening
$\Gamma_{p}=\omega_{E}/5$. This corresponds to the Debye temperature
\begin{figure}[h]
\includegraphics[width=0.6\columnwidth]{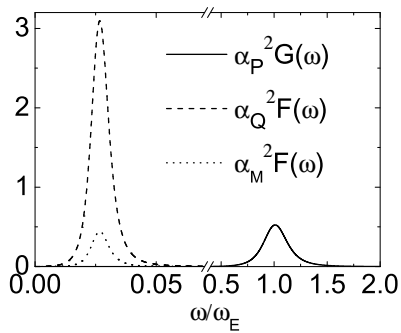}
\caption{Calculated bosonic Eliashberg function $H_{\pm}(\Omega)$
for La$_{1-x}$Pr$_{x}$Os$_{4}$Sb$_{12}$. The solid curve denotes the
phonon contribution while the dashed and dotted curves refer to
the quadrupolar and magnetic contributions, respectively.}
\label{fig2}
\end{figure}
of this compound \cite{Vollmer} (see Fig.~\ref{fig2}). Setting $\mu^*
= 0.1$ and $\lambda = 2 \int_{-\infty}^{+\infty} d\Omega
\alpha_{P}^{2}G(\Omega)/ \Omega = $ 0.33 the Eliashberg
equations yield $T_c =0.74$ K and $2 \Delta_0 / k_B T_c\approx $
3.5. In Fig.~\ref {fig3}(a) we show the calculated results for the
real part of $\Delta(\omega)$ and the renormalization function
$Z(\omega)$.
\begin{figure}[h]
\includegraphics[width=0.6\columnwidth]{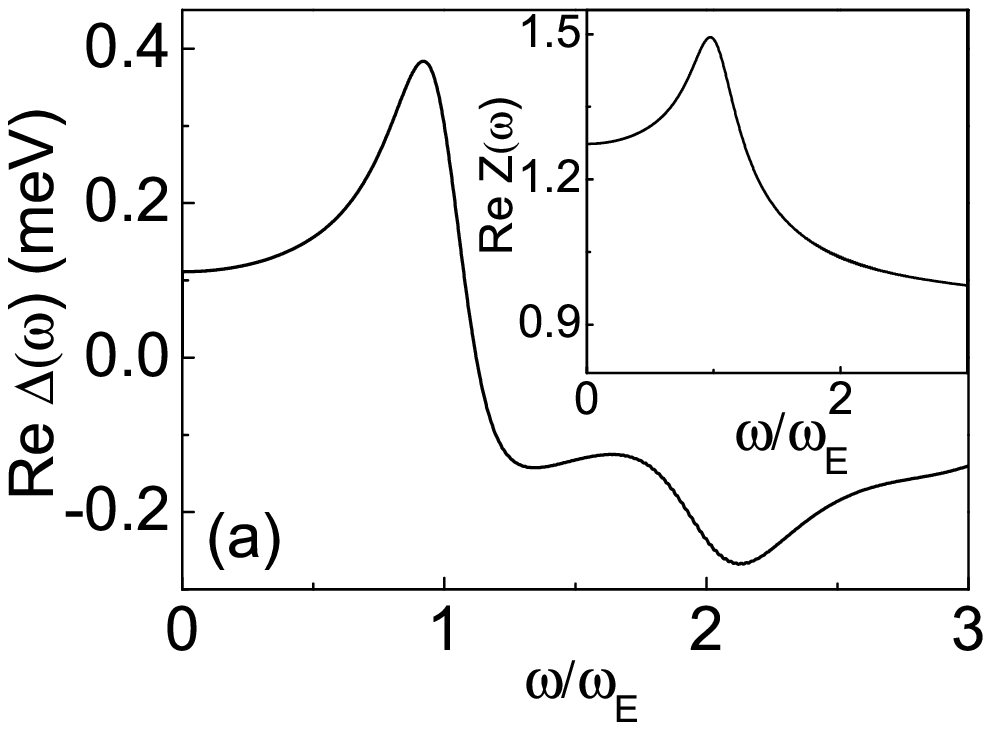}
\includegraphics[width=0.6\columnwidth]{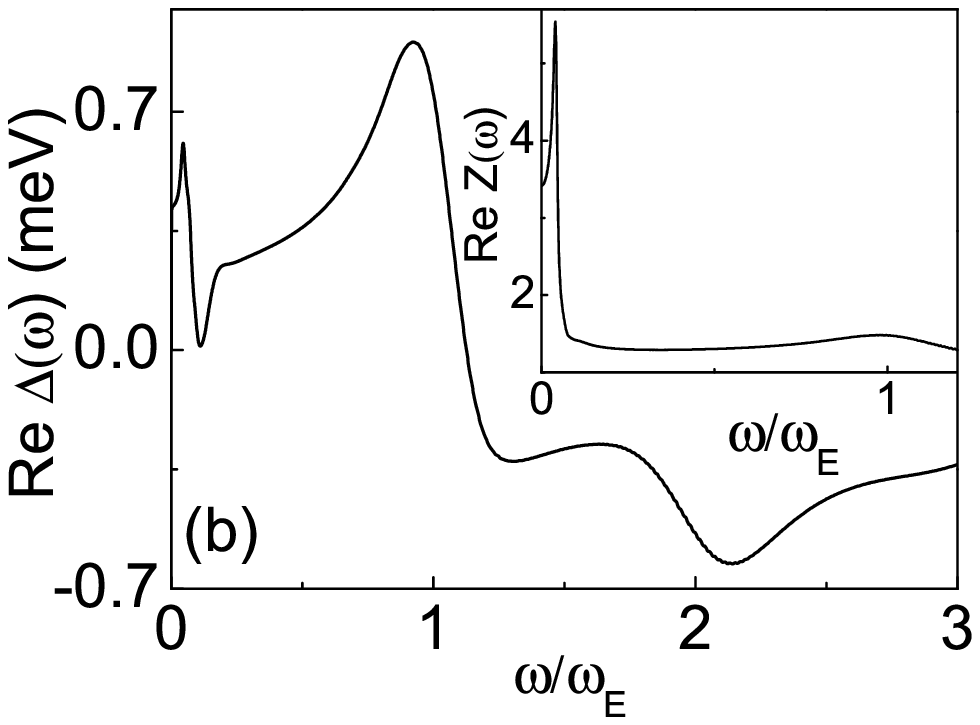}
\caption{Calculated frequency dependence of the real part of the
superconducting gap function $\Delta(\omega)$ for
LaOs$_{4}$Sb$_{12}$ (a) and PrOs$_{4}$Sb$_{12}$ (b) compounds. The
renormalization function, $Z(\omega)$ at $T=0.04$ K is shown in the
insets. Note that we set the cutoff frequency equal to $5\omega_{E}$ and
introduce a finite damping $\Gamma = 0.01$ meV.} \label{fig3}
\end{figure}
In both cases one finds typical behavior of a BCS-like
superconductor. In particular, Re$\Delta(\omega)$ shows a peak
structure at $\omega_E$ and becomes negative at larger frequencies, 
reflecting the effective repulsion for $\omega >\omega_E$. For
$\omega_E >> \Delta_0$, the renormalization function is close
to 1 and the ratio $2\Delta_0 / k_B T_c$ equals the BCS value.

A similar crystallographic structure and nearly equal ionic radii of
the La and Pr ions allow to assume nearly the same phonon modes in
LaOs$_{4}$Sb$_{12}$  and PrOs$_{4}$Sb$_{12}$. The main difference is
the occurrence of the low-energy exciton in PrOs$_{4}$Sb$_{12}$ at
about $\Delta_{CEF} \approx 0.03 \omega_E$. As discussed above there
are both (magnetic) pair-breaking and (quadrupolar) pair-forming
scattering processes of electrons by the excitons. Both processes
can be taken into account as separate contributions to the
Eliashberg function $\alpha^2 F(\omega)$. Using Eq.~(\ref{exciton})
and ignoring a weak dispersion of the exciton as observed by
inelastic neutron scattering (INS) experiments\cite{quadr2} we obtain, 
after averaging over the FS, $\lambda_{Q} = 2 \int_{0}^{\infty}
d\omega \alpha_{Q}^{2}F(\omega)/\omega \approx I_{ac}^{2}
|M_{14}^{Q}|^{2} 2N_{0}/\Delta_{CEF}$ where
$|M_{14}^{Q}|^{2}=35(1-d^{2})$ is the quadrupolar scattering matrix
element and $N_{0} = 7.625$ eV$^{-1}$ is the bare (unrenormalized)
electronic density of states.\cite{Vollmer} Furthermore, we set
$I_{ac}\approx 2$ meV. Similarly, we find $\lambda_{M} = 2
\int_{0}^{\infty} d\omega \alpha_{M}^{2} F(\omega)/\omega
\approx (1-g_{L})^{2} I_{ex}^{2} |M_{14}^{M}|^{2}
2N_{0}/\Delta_{CEF}$ where, $M_{14}^{M}=20d^{2}/3$
is the magnetic scattering matrix element and $g_{L}=$0.8 is the
Land$e'$ factor. Following our previous estimates, we set
$\alpha_{Q}^{2}/\alpha_{M}^{2} = 7$ and obtain $I_{ex} \approx 16.4
I_{ac}$, where $I_{ex}$ is the on-site magnetic interaction constant
between exciton and conduction electrons. All components entering
the total Eliashberg function $H_{\pm}(\Omega)$ are shown in Fig.
~\ref{fig2}. As expected the quadrupolar contribution is about 7
times larger than the magnetic one. Therefore an inclusion of the
excitons yields an enhancement of the superconductivity in
PrOs$_{4}$Sb$_{12}$ as compared to LaOs$_{4}$Sb$_{12}$. In
particular, the total value of $\lambda$ increases to 3.05 yielding
strong coupling superconductivity in PrOs$_{4}$Sb$_{12}$ with $T_c =
1.85$ K. Here, one has to remember that in PrOs$_4$Sb$_{12}$ there
are three contributions to the total $\lambda$ and their sum gives
$\lambda$=3.05. This yields a strong renormalization of the
quasiparticle mass in PrOs$_4$Sb$_{12}$ compared to its
LaOs$_4$Sb$_{12}$ counterpart. At the same time $T_c$ and
$\Delta(\omega)$ are determined by  Eq.~(5) where the difference
between quadrupolar and magnetic scattering enters. Therefore the
enhancement of $T_c$ is moderate despite the large change in
$\lambda$.

\begin{figure}[t]
\includegraphics[width=0.6\columnwidth]{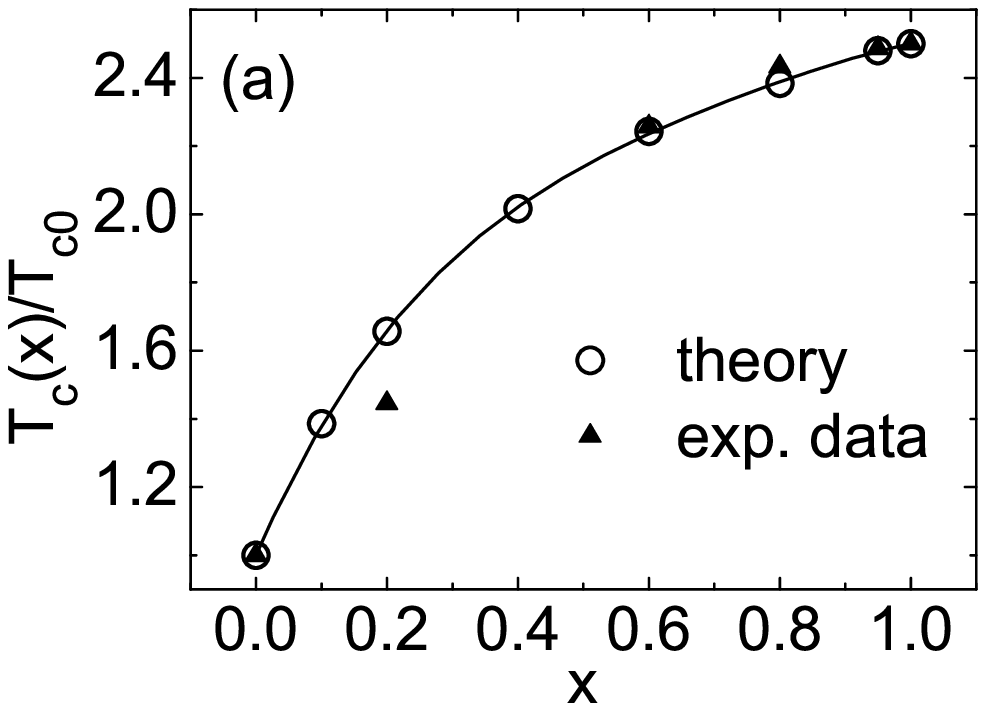}
\includegraphics[width=0.6\columnwidth]{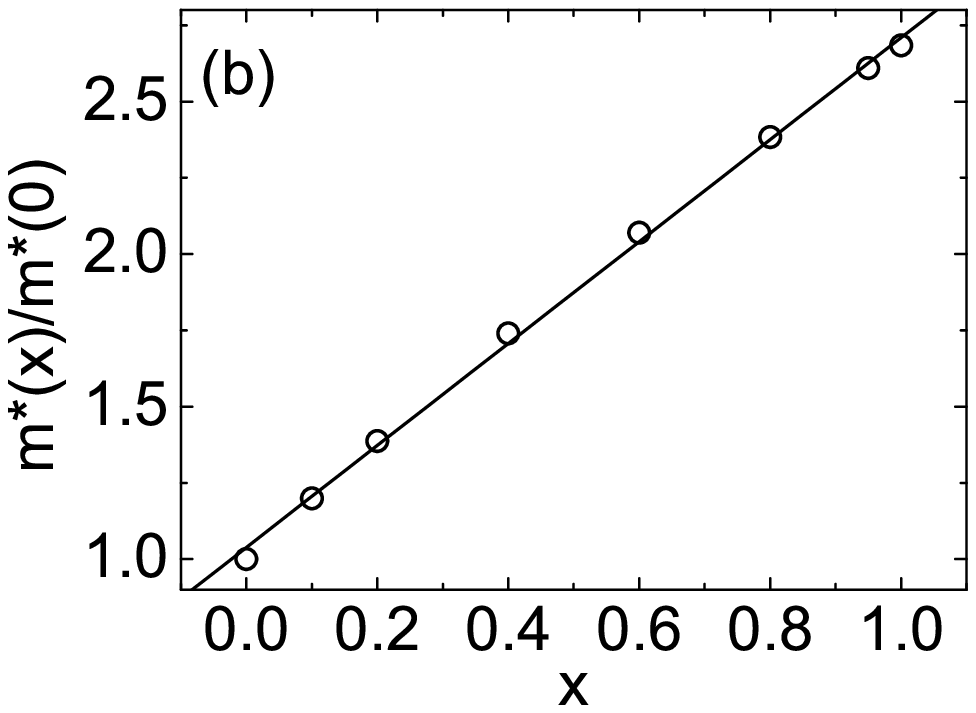}
\caption{Calculated superconducting transition temperature $T_c$ (a)
and quasiparticle effective mass $m^*$(b) for
La$_{1-x}$Pr$_{x}$Os$_{4}$Sb$_{12}$ as a function of Pr
concentration $x$. The straight curve is a guide to the eye. The
experimental data are taken from Ref.~\protect\onlinecite{Yogi}.}
\label{fig5}
\end{figure}

Figure \ref{fig3}(b) shows the calculated frequency dependence of
the real part of the superconducting gap and the renormalization
function for PrOs$_{4}$Sb$_{12}$. One notices a sharp feature at an
energy of about $\Delta_{CEF} + \Delta_0$ and a strong
renormalization of the quasiparticle mass due to the presence of the
low-energy exciton. It is interesting to note that both magnetic and
quadrupolar scatterings contribute to the renormalization of the
quasiparticle mass while only their difference contributes to
Cooper-pair formation. We suggest that the low-energy peak in
$\Delta(\omega)$ should be visible in the tunneling density of
states which could be checked experimentally. Another remarkable
feature is that we find $2\Delta_0/k_B T_c \approx $5.4 typical for
a strong-coupling superconductor. Therefore, within this approach we
find PrOs$_{4}$Sb$_{12}$ to be a conventional $s$-wave
strong-coupling superconductor, with enhanced $T_c$ due to
low-energy excitons with mainly quadrupolar scattering. It has been
proposed previously that quadrupolar scattering may alone yield
unconventional superconductivity in PrOs$_{4}$Sb$_{12}$ due to the
exciton dispersion.\cite{quadr1,quadr2} Indeed such a possibility
cannot be excluded from our calculation. At the same time we find it
difficult to reconcile with the experimental evolution of $T_c$ in
La$_{1-x}$Pr$_{x}$Os$_{4}$Sb$_{12}$ as a function $x$. In
particular, assuming the linear dependence of the exciton's
contribution to $H_{\pm}(\omega)$ as a function of $Pr$
concentration we find good agreement between calculated and measured
$T_c (x)$ [see Fig.~\ref{fig5}(a)]. At the same time, unconventional
superconductivity in PrOs$_{4}$Sb$_{12}$ would be unstable with
respect to nonmagnetic La impurities and $T_c$ should vanish at a
small concentration of La. In Fig.~\ref{fig5} (b) we show the
evolution of the effective mass as a function of Pr concentration.
We find that the effective mass increases by a factor of 2.5 from
LaOs$_{4}$Sb$_{12}$ to PrOs$_{4}$Sb$_{12}$ which agrees with dHvA
data.\cite{sugawaraF} An inclusion of the higher-lying CEF energy
levels provides stronger renormalization of the effective
mass.\cite{zwicknagl}
It is interesting to note that experimentally $\Delta C/ T_c$ shows
a nonlinear dependence on the La concentration, in particular 
close to PrOs$_4$Sb$_{12}$\cite{Rotundu} which cannot be explained by
our theory. Whether this is a sign of the first-order phase
transition between different symmetries of the superconducting order
parameter or is a result of the Pr-Pr interaction neglected in our
theory remains to be understood both theoretically and
experimentally. Finally, let us note that our theory allows us also to
explain qualitatively the behavior of $T_c$ in
Pr(Os$_{1-x}$Ru$_x$)$_4$Sb$_{12}$ for $x <$ 0.6. In particular, it
has been found that the CEF level splitting between the lowest
levels increases as a function of $x$ while $T_c$
decreases.\cite{frederick} According to our Fig.~\ref{fig1}(a) the
Cooper-pair constructive quadrupolar scattering decreases with
increasing $x$ while the magnetic pair-breaking contribution tends to
saturate. Therefore, the $T_c$ should decrease as it is also
observed in the experiment.\cite{frederick,remark1}

In conclusion by solving the nonlinear Eliashberg equations we find
that in La$_{1-x}$Pr$_{x}$Os$_{4}$Sb$_{12}$ the dominant quadrupolar
scattering due to the specific $T_h$ crystal group symmetry of the
lattice is responsible for the observed increase of $T_c$ as a
function of Pr concentration. Our analysis suggests that a
combination of conventional electron-phonon interaction together
with pairing mediated by quadrupolar excitons yields strong-coupling
superconductivity in the PrOs$_{4}$Sb$_{12}$ system.

We are thankful to G. Zwicknagl, A. Yaresko, S. Burdin, and T.
Westerkamp for helpful discussions.


\begin{thebibliography}{11}
\bibitem{remark1} One should note,
however, that the substitution of Os by Ru results not only in the
change of CEF effects but also in the change of spin-orbit coupling
strength within the conduction band and also different electronic
correlations.
\end{thebibliography}
\end{document}